# Introduction of a Triple Prime Symmetric Key Block Cipher

Abhijit Chowdhury
NSHM College of Management
& Technology, Durgapur
West Bengal, INDIA

Angshu Kumar Sinha
NSHM College of Management
& Technology, Durgapur
West Bengal, INDIA

Saurabh Dutta
Dr. B.C Roy Engineering
College
West Bengal, INDIA

## ABSTRACT
This paper proposes to put forward an innovative algorithm for symmetric key block cipher named as "Triple Prime Symmetric Key Block Cipher with Variable Key-Spaces (**TPSKBCVK**)" that employs triple prime integers as private key-spaces of varying lengths to encrypt data files. Principles of modular arithmetic have been elegantly used in the proposed idea of the cipher. Depending on observations of the results of implementation of the proposed cipher on a set of real data files of several types, all results are registered and analyzed. The strength of the underlying design of the cipher and the liberty of using a long key-space expectedly makes it reasonably non-susceptible against possible cryptanalytic intrusions. As a future scope of the work, it is intended to formulate and employ an improved scheme that will use a carrier media (image or multimedia data file) for a secure transmission of the private keys.

## 1. INTRODUCTION
Many symmetric key cipher protocols are developed with application of Modular Arithmetic [1]. Principles of Modular Arithmetic have been applied elegantly to develop the proposed Triple Prime Symmetric Key Block Cipher [2] [3] with the acronym **TPSKBCVK.** The encryption of data using TPSKBCVK results in encrypted data with expanded length, but the strength of the underlying design of the cipher and the liberty of using long varying length key space can make it fairly non- susceptible against probable cryptanalytic intrusions.

Section 2 introduces the scheme of the proposed cipher TPSKBCVK. An account of results observed on implementation of TPSKBCVK on a sample plaintext is demonstrated in section 3. Results obtained from a set of real implementations are exhibited in section 4. Section 5 is an analytical study of TPSKBCVK. A conclusion is drawn in section 6.

## 2. THE SCHEME
The source data targeted to be encrypted is read as blocks of B-Bits. All such B-bit blocks generated are converted to Base-10 integers. Sufficiently large triple prime integers KEY1, KEY2, KEY3 are taken as private keys satisfying the criteria:

1. Product of KEY1 and KEY2 has to be sufficiently large than the integers to be encrypted.

Sections 2.1 and section 2.2 respectively explains the encryption and the decryption scheme with conventionally *Alice* considered as the sender and *Bob* considered as the receiver.

## 2.1 Encryption
The steps of computations required to be followed in the encryption end for P being a positive integer to be encrypted are as follows:

1. Set N=$(KEY1 \times KEY2)^2$
2. Set K3 = $\varphi$ (KEY1 x KEY2) – 1
3. Set B = $(P)^{K3}$ mod N
4. Set C = $(B \times KEY3)^{K3}$ mod N

The function $\varphi$ (KEY1 x KEY2) is known as a totient function. The totient $\varphi$ (K) of a positive integer K defines the number of integers less than or equal to K that are relatively prime to K. $\Phi$ (K) = K – 1, where K is a prime number.

## 2.2 Decryption
The set of values for KEY1, KEY2 and KEY3 are extracted from the private key, the following steps of operations are carried out in the decryption end.

1. Set N=(KEY1 x KEY2)
2. Set K3 = $\varphi$ (KEY1 x KEY2) – 1
3. Set B = $(C \times KEY3)^{K3}$ mod N
4. Set P = $B^{K3}$ mod N

The function $\varphi$ (KEY1 x KEY2) is a totient function as illustrated in section 2.1[4].

## 3. IMPLEMENTATION
This section demonstrates the steps of encryption and decryption using a small test plaintext WORLD considering KEY1 = 17, KEY2 = 19 and KEY3=23.

Table 1 registers different transitional result and the final result of implementation following the encryption algorithm explained in section 2.1. Finally the cipher text þŽ☐ ž ` Ù¥ "( Áµ is derived for the taken plaintext.

**Table 1 :** Implementation of Encryption of the Plaintext "WORLD"

| CHARACTER | ASCII VALUE | Value obtained by following step 3 of the Encryption Algorithm | Value obtained by following step 4 of the Encryption Algorithm | Printable Character (including spaces) corresponding to value obtained through step 4 of the Encryption Algorithm |
|---|---|---|---|---|
| W | 87 | 92404 | 102142 | þŽ |
| O | 79 | 14396 | 24734 | ž` |
| R | 82 | 36306 | 42457 | Ù¥ |
| L | 76 | 39710 | 75810 | "( |
| D | 68 | 4624 | 46529 | Áµ |

A reverse process is followed to implement the decryption algorithm stated in section 2.2.

## 4. RESULTS

A viable implementation of the encryption and decryption algorithm established in section 2.1 and section 2.2 respectively was performed using C language and MIRACL (**M**ulti precision **I**nteger and **R**ational **A**rithmetic **C**/C++ **L**ibrary) [5] for big integer computations. The plaintext was divided and read as blocks of length 8-bits each; a set of triple prime integers KEY1, KEY2 and KEY3 each 8-bit long, was considered as private keys for operation on the plaintext data. Generated Ciphertext was written to file as blocks of 32-bits to get around overflow problem, as in few cases 8-bit data after encryption is to produce values larger than 255. File types .txt, .doc, .docx, .xls, .pdf, .mp4, .ocx, .jpg, .bmp, .dll, .gif, .sys etc were encrypted and effectively decrypted.

A comparative study of file type, source file size, decrypted file size, encryption time and decryption time is illustrated in table 2 and table 3.

**TABLE 2**:File type, source file size, encryption time, encrypted file size

| filename | size (in KB) | encryption time (in sec) | encrypted file size (in KB) |
|---|---|---|---|
| Abc.docx | 47 | 1.625 | 188 |
| Ap_hawaii.jpg | 18 | 0.688 | 71 |
| ENG1.docx | 47 | 1.609 | 104 |
| Cv.doc | 60 | 1.328 | 238 |
| Effectctrl.ocx | 405 | 11.906 | 1617 |
| Fishing.bmp | 17 | 0.687 | 68 |
| Hmmapi.dll | 59 | 1.625 | 236 |
| Marks.xls | 85 | 1.797 | 338 |
| Project2.zip | 66 | 2.375 | 261 |
| Test.txt | 2 | 0.125 | 7 |
| Tabulator.pdf | 434 | 14.656 | 1736 |
| **TABLE 2** continued: File type, source file size, encryption time, encrypted file size ||||
| Quotes.gif | 25 | 0.953 | 100 |
| Bee.wav | 144 | 5.172 | 573 |
| Clock.avi | 81 | 2.516 | 324 |
| Greenstone.bmp | 26 | 0.828 | 104 |
| Handsafe.reg | 1 | 0.063 | 3 |
| Himem.sys | 5 | 0.234 | 19 |
| IE7Eula.rtf | 73 | 2.687 | 292 |
| Jview.exe | 151 | 3.281 | 602 |
| Keyboard.drv | 2 | 0.109 | 8 |
| Mscomm32.ocx | 102 | 3 | 406 |
| Normnfd.nls | 39 | 1.094 | 154 |
| Notepad.exe | 68 | 1.891 | 270 |
| Preconvertlite.dll | 103 | 3.14 | 412 |
| Setup.xml | 4 | 0.234 | 16 |

**TABLE 3**: Encrypted File type, encrypted file size, and decryption time

| Encrypted File name | Encrypted file size (in KB) | Decryption time (in sec) |
|---|---|---|
| Abc.docx.enc | 188 | 1.625 |
| Ap_hawaii.jpg.enc | 71 | 0.703 |
| ENG1.docx.enc | 104 | 1.625 |
| Cv.doc.enc | 238 | 1.265 |
| EffectCtrl.ocx.enc | 1617 | 11.984 |
| Fishing.bmp.enc | 68 | 0.672 |
| Hmmapi.dll.enc | 236 | 1.609 |
| Marks.xls.enc | 338 | 1.781 |
| Project2.zip.enc | 261 | 2.359 |
| Test.txt.enc | 7 | 0.125 |
| Tabulator.pdf.enc | 1736 | 14.734 |
| Quotes.gif.enc | 100 | 0.953 |
| Bee.wav.enc | 573 | 5.172 |
| Clock.avi.enc | 324 | 2.531 |
| Greenstone.bmp.enc | 104 | 0.906 |
| Handsafe.reg.enc | 3 | 0.047 |
| Himem.sys.enc | 19 | 0.234 |
| IE7Eula.rtf.enc | 292 | 2.656 |
| Jview.exe.enc | 602 | 3.25 |
| Keyboard.drv.enc | 8 | 0.11 |
| Mscomm32.ocx.enc | 406 | 2.953 |
| Normnfd.nls.enc | 154 | 1.109 |
| Notepad.exe.enc | 270 | 1.859 |
| PreConvertLite.dll.enc | 412 | 3.141 |
| Setup.xml.enc | 16 | 0.234 |

| TABLE 4: Average Required Time for Exhaustive Key Search | | | |
|---|---|---|---|
| Key Size (Bits) | Number of Alternative Keys | Time Required at 1 Encryption /µs | Time Required at $10^6$ encryptions /µs |
| 56 | $2^{56} = 7.2 \times 10^{16}$ | $2^{55}$ µs =1142 years | 10.01 hours |
| 128 | $2^{128} = 3.4 \times 10^{38}$ | $5.4 \times 10^{24}$ years | $5.4 \times 10^{18}$ years |
| 168 | $2^{168} = 3.7 \times 10^{50}$ | $2^{167}$ µs = $5.9 \times 10^{36}$ years | $5.9 \times 10^{30}$ years |
| 26 characters | $26! = 4 \times 10^{26}$ | $2 \times 10^{26}$ µs = $6.4 \times 10^{12}$ years | $6.4 \times 10^6$ years |

It can be observed from data in Table 2 and Table 3 that the relation between encryption time and file size is linear. The average time to encrypt 1 KB of data is calculated to be 0.030825097 seconds. Also, it was found that the average decryption time of 1 KB of encrypted data is 0.007811096 seconds.

Encrypted data produced from 8-bit source data is written as 32-bit data, so the encrypted data file size (in bytes) is observed to be about four times the source data file size (in bytes). Thus it was observed that the proposed cipher accompanies an overhead of data expansion. An operational overhead of being marginally slow in encryption is observed in its practical implementation.

## 5. ANALYSIS

Small keys were used to exhibit results of encryption and decryption using TPSKBCVK though much larger key-spaces can be used in practice.

The effective length of key is equal to product of length of KEY1 and KEY2. Brute-force attack to the proposed cipher scheme will require searching entire search space of $2^{\text{(product of length of KEY1 \& KEY2)}}$. The projected mean time required for such extensive search is given in table 4[6].

The proposed cipher uses exponentiation hence data expansion in encryption process is expected. Encrypted file size can be reduced significantly by writing encrypted data as bit stream and formulating appropriate padding schemes to mark separate blocks. Faster methods of calculating large integer powers of integers can be adopted to reduce both encryption and decryption time. Encryption process can be optimized to be faster if read block size (in bits) is increased but it should be less than the size of key (in bits). The proposed scheme restricts use of the read block size greater than the product of KEY1 and KEY2.

## 6. CONCLUSION

The design of proposed cipher TPSKBCVK makes it extensible for implementation with larger keys. Time required for encryption and decryption using TPSKBCVK is almost equal. If brute-force attack is employed to identify the three 128-bit prime integer keys used as keys for encryption then at least all the primes in the range of integers from 1 to $2^{128}$ are to be tested three times to find all the three probable keys. The approximate number of primes calculated from the relation $\pi(n) \approx n/\ln(n)$, where $1 <= n <= 2^{128}$ is found to be $383534127545935 \times 10^{22}$. It is observed that if one trillion of these numbers per second can be checked then more than $121{,}617{,}874{,}031{,}562 \times 10^3$ years is needed to test out all such primes. The requirement of time as shown in table 4, to calculate the keys makes it reasonably non- susceptible against cryptanalytic attacks [7]. As a future scope of the work, it is intended to formulate and employ an improved scheme that will use a carrier media (image or multimedia data file) for a secure transmission of the private keys.